\documentclass[aps,prl,twocolumn,floatfix,showpacs,amssymb]{revtex4-1}
\usepackage[utf8]{inputenc}
\usepackage[english]{babel}
\usepackage{amsmath}
\usepackage{amsfonts}
\usepackage{amssymb}
\usepackage{graphicx}
\usepackage{float}
\usepackage{color}
\usepackage{subfigure}
\usepackage{braket}

\begin{document}

\title{Multicomponent electron-hole superfluidity and the BCS-BEC crossover \\in double bilayer graphene}
\author{S. Conti$^{1,2}$, A. Perali$^{1}$, F. M. Peeters$^{2}$, and D. Neilson$^{1,2}$}
\affiliation{$^1$Dipartimenti di Fisica e di Farmacia, Universit\`a di Camerino, 62032 Camerino (MC), Italy\\
$^2$Department of Physics, University of Antwerp, Groenenborgerlaan 171, B-2020 Antwerpen, Belgium}

\begin{abstract}
Superfluidity in coupled electron-hole sheets of bilayer graphene is predicted here to be multicomponent because of the conduction and valence bands.
We investigate the superfluid crossover properties as functions of the tunable carrier densities and  the tunable energy band gap $E_g$.
For small band gaps there is a significant boost in the two superfluid gaps, but the interaction driven excitations from the valence  to the conduction band can  weaken the superfluidity, even blocking the system from entering the BEC regime at  low densities.  At a given larger density, a  band gap  $E_g\sim 40$-$60$ meV  
can carry the system into the  strong-pairing multiband BCS-BEC crossover regime, the optimal range for realization of high-$T_c$ superfluidity.
\end{abstract}
\pacs{
71.35.-y , 
73.21.-b, 
73.22.Gk 
74.78.Fk 
}
\maketitle

The recent fabrication of two very close, but electrically isolated, conducting bilayer graphene sheets, one containing electrons and the other holes
\cite{geim_vanderwaals,lee_giantDrag,li_negativeDrag}, raises exciting possibilities of observing high-temperature superfluidity\cite{perali_dbg}, 
since the electrons form pairs with the holes through very strong Coulomb attraction \cite{lozovik_superc, shevchenko_theory}.

In bilayer graphene, the Fermi energy can be tuned continuously relative to the average strength of the Coulomb interactions between carriers \cite{zarenia_enhancement}. 
Metal gates can be used to change the carrier densities\citep{lee_chemical} so as to tune each sheet from the high-density regime of weak interactions,
 to the low-density regime where the average Coulomb interactions between carriers are much larger than their kinetic energies.  
However, the touching of the conduction and valence bands at the semi-metallic point means  that
at low densities, carriers from the two bands can strongly  affect each other, and this weakens the superfluid pairing.
A tunable energy band gap inserted between the conduction and valence bands by application of electric fields perpendicular to the sheets \cite{zhang_bandgap}, can be used to decouple the conduction and valence bands.
By contrast, in multicomponent high-$T_c$ iron-based superconductors, the carrier densities are difficult to tune and their energy band structure 
is fixed \cite{bianconi2013quantum, rinott2017tuning}.

We investigate the effect of the multibands  at zero temperature on the superfluid BCS-BEC crossover and  BEC regimes as functions of  
the tunable carrier densities and energy band gap $E_g$, and we identify an optimal combination of the experimental parameters for superfluidity.  
We find that the crossover properties depend sensitively on both the carrier densities and the band gap.


Our effective Hamiltonian is,\\

\ \\
\vspace*{-1.8em}
\begin{equation}
\begin{aligned}
H=&\sum_{k\gamma} \;\left\{\xi^{(e)\gamma}_{k} \,c^{\gamma\dagger}_{k} \, c^{\gamma}_{k} +
\xi^{(h)\gamma}_{k} \,d^{\gamma\dagger}_{k} \, d^{\gamma}_{k} \right\} \\ 
 +& \sum_{k,k',q,\gamma,\gamma'} V_{k\, k'}
\,c^{\gamma\dagger}_{k+ q/2}
\,d^{\gamma\dagger}_{-k+ q/2}
\, d^{\gamma'}_{-k'+ q/2}
\, c^{\gamma'}_{k'+ q/2}\ .
\end{aligned}
\label{eq:Hamiltonian}
\end{equation}
The band index $\gamma=\pm$ labels the conduction and valence bands of each bilayer sheet. 
$c^{\gamma\dagger}_{k}$ and $d^{\gamma\dagger}_{k}$  are the creation operators in band $\gamma$ for the  electrons and holes 
in their respective bilayer sheets, and $c^{\gamma}_{k}$  and $d^{\gamma}_{k}$  the corresponding destruction operators.  
Spin indices are  implicit.  We make the standard transformation so the bands of the p-doped bilayer are filled with positively charged holes 
up to the Fermi level located in the conduction band. 
$V_{k\, k'}$ is the electron-hole interaction.
$\xi^{(e,h)\gamma}_{k}=\varepsilon^{\gamma}_k -\mu$, where we take the single-particle energy  dispersions of the conduction and valence bands for each bilayer graphene sheet $\varepsilon^{\gamma}_k$ to be identical and parabolic: $\varepsilon^{+}_k= \hbar^2 k^2/2m^*$ and $\varepsilon^{-}_k= -\hbar^2 k^2/2m^* -E_g$. 
We take the effective mass for electrons and holes equal, $m^*=m^\star_e=m^\star_h=0.04m_e$ \cite{zou_effectivem}.
We set the chemical potential $\mu$ equal in the two bilayer sheets, considering only equal electron and hole densities.
  
We consider intraband pairing and Josephson-like pair transfer between the conduction and valence bands. 
The neglect of crosspairing will be justified later in the paper.
The coupled zero temperature gap equations are \cite{lozovik_multiband}, 
\begin{equation}
\Delta^{\gamma}_k=-\sum_{k',\gamma'} F^{\gamma\gamma'}_{kk'} \; V_{k\, k'} \; \frac{\Delta^{\gamma'}_{k'}}{2 E^{\gamma'}_{k'}}\ .
\label{eq:gap}
\end{equation}
\vspace*{-1.2em}
\begin{equation}
\begin{aligned}
E^{\gamma}_{k}=&\sqrt{(\xi^{\gamma}_{k})^2 + (\Delta^{\gamma}_{k})^2} \qquad 
\xi^{\gamma}_{k}     =\frac{1}{2}\left(\xi^{(e)\gamma}_{k}+ \xi^{(h)\gamma}_{k}\right) \\[1.0ex]
(v^\gamma_{k})^2 =& \frac{1}{2}\left(1 -\frac{\xi^{\gamma}_{k}}{E^{\gamma}_{k}} \right) \qquad
(u^\gamma_{k})^2 = \frac{1}{2}\left(1 + \frac{\xi^{\gamma}_{k}}{E^{\gamma}_{k}} \right)\ .\\[1.0ex]
\label{eq:B_A}
\end{aligned}
\end{equation}
\vspace*{-1.2em}
\begin{small}
\begin{equation}
F^{\gamma\gamma'}_{\mathbf{kk'}}=\frac{1}{2}[1+ \gamma\gamma' (\cos\alpha_k\cos\alpha_{k'} +\sin\alpha_k\sin\alpha_{k'} \cos2\phi) ]
\label{eq:formfactor}
\end{equation}
\end{small}
\!\!\!\!\!
is the  form factor for the overlap of the single-particle state $\Ket{k}$ in band $\gamma$ with $\Ket{k'}$ in band $\gamma'$,  
$\phi= \cos^{-1}(\widehat{\mathbf{kk'}})$, and $\alpha_k= \tan^{-1}\left\{\hbar^2 k^2 /( m^* E_g)\right\}$ \cite{wang_coulomb}. 
We note the dependence of $F^{\gamma\gamma'}_{\mathbf{kk'}}$ on $E_g$.

To determine the chemical potential $\mu$, we take for the density control parameter for each bilayer sheet \cite{barlas_chirality, hwang_coulomb},
\begin{equation}
 n^+_0= g_s g_v \sum_k \left[(v^+_k)^2 - (u^-_k)^2\right] \ .
 \label{eq:density}
\end{equation} 
$g_s=g_v=2$ are the spin and valley degeneracy for bilayer graphene.
$\mu$ is then obtained by solving Eqs.\ \ref{eq:gap} to \ref{eq:density}. 
$n^+_0$ is defined as the total number of carriers in the conduction band, $n^+= g_s g_v \sum_k (v^+_k)^2$, 
less the number of carriers in the conduction band that have been excited from the valence band. 
The number of such excited carriers in the conduction band equals the number of unoccupied states left behind in the valence band, $ g_s g_v \sum_k (u^-_k)^2$. 

The reason for this choice of control parameter (Eq.\ \ref{eq:density})  is due to the influence of the valence band on the conduction band. 
The presence of the valence band means that the overall number of carriers in the conduction band, $n^+$, is no longer controlled purely by doping or using the metal gates, 
as is the case for the single band system, since now there are additional carriers in the conduction band excited from the valence band due to interactions. 
This increase in the number of carriers in the conduction band will push up the Fermi energy.  
We use $n^+$ to define an effective Fermi momentum $k^*_F= \sqrt{4\pi n^+/g_sg_v}$, and effective Fermi energy in the conduction band $E^*_F=(\hbar k^*_F)^2/2m^*$.

For large $n^+_0$, the average kinetic energy of the carriers in the conduction band $\langle K\rangle$ is large relative to the average strength of the Coulomb interactions 
$\langle V\rangle$, and, since $\sum_k (v^+_k)^2\gg \sum_k (u^-_k)^2$, there are only a negligible number of carriers excited out of the valence band.
However, small $n^+_0$ does not necessarily imply that $\langle K\rangle \ll \langle V\rangle$, since for sufficiently small $E_g$, 
both $ \sum_k (v^+_k)^2$ and $ \sum_k (u^-_k)^2$ can be large but nearly equal. 
We will see that both $n^+_0$ and $E_g$ play important roles in determining the relative strength of the Coulomb interactions.  

We take the interaction term in Eq.\ \ref{eq:gap} as unscreened, 
\begin{equation}
V_{k\, k'}=-\frac{2\pi e^2}{\epsilon}\frac{e^{-d|\textbf{k}-\textbf{k}'|}}{|\textbf{k}-\textbf{k}'|}\ ,
\end{equation}
where $d$ is the thickness of the insulating barrier separating the two bilayer sheets.  A hexagonal Boron Nitride insulating barrier with dielectric constant $\epsilon=3$ 
and thickness $d\geq1$ nm can electrically isolate the two bilayer sheets
\cite*{dean_boron, kim_hbn}.  We set $d=1$ nm.   

Neglecting screening is an excellent approximation in the BEC regime where the strong interactions tightly bind the pairs, making them compact on the scale of the average inter-carrier separations $r_0$ \cite{neilson_excitonic,*Maezono2013}.  
For example, at a carrier density of $1\times 10^{11}$ cm$^{-2}$, $r_0=18$ nm which is much larger than our $d$.  
In this case, the superfluid gap in the excitation spectrum is large on the scale of $E^*_F$, 
and this suppresses the low energy excitations responsible for screening \cite{lozovik_condensation}.   
The unscreened approximation continues to be remarkably 
good even in the BCS-BEC crossover regime at intermediate densities \cite{perali_dbg}, predicting superfluid gaps correctly to within $\sim 20\%$ \cite{neilson_excitonic}.  
However, at larger densities, $n^+_0 \agt 5 \times 10^{11}$ cm$^{-2}$, the unscreened approximation is known to completely break down, since at such densities, 
onset of very strong screening 
completely suppresses superfluidity in what would otherwise have been the BCS regime \cite{lozovik_condensation}. 
For this reason we will restrict our  results to densities $n^+_0 \leq 5 \times 10^{11}$ cm$^{-2}$.   

We omit intralayer electron-electron and hole-hole interactions. 
This approximation can be justified by comparing the gaps calculated including correlations between like-species\cite{zhu_excitonmodel}, 
with the gaps calculated neglecting these correlations\cite{pieri_imbalance}.
The intralayer correlations have at most a $10$-$20$\% effect on the superfluid gap. 

In general, the regimes of the crossover phenomena in a one-band system are conveniently characterized by the superfluid condensate fraction $c$ \cite{giorgini_condensate,*Salasnich_condensate,*Guidini2014}.  
$c$ is defined as the fraction of carriers bound in pairs relative to the total number of carriers.  
The usual classification is: for $c>0.8$ the condensate is in the BEC regime, for $c<0.2$ in the BCS regime, and otherwise in the crossover regime.

However, we have here two partial condensate fractions, $c^{\pm}$, for the conduction and valence bands.
For the conduction band the usual one-band expression 
is readily generalized to the number of pairs divided by the total number of carriers in the conduction band,
\begin{equation}
c^{+}=\frac{\sum_{k} (u^+_k)^2\; (v^+_k)^2}{\sum_{k} (v^+_k)^2}\ ,
\label{eq:c+}
\end{equation}
but for the valence band the corresponding definition of $c^-$ is the ratio of the number of pairs in the valence band to the number of anti-particles in the valence band,
\begin{equation} 
c^{-}=\frac{\sum_{k} (u^-_k)^2\; (v^-_k)^2}{ \sum_{k} (u^-_k)^2}\ .
\label{eq:c-}
\end{equation}
(We use the term {\it anti-particle} to refer to an empty single-particle state in the valence band, since we reserve the term {\it hole} to refer to the hole-doped bilayer sheet.)
At zero temperature, the valence band anti-particles are generated exclusively as a result of the effect of interactions that excite carriers out of the valence band 
up into the conduction band. The pairs in the valence band are formed from pairing of anti-particles 
of the two sheets. 


\begin{figure}[h]
\includegraphics[width=\columnwidth]{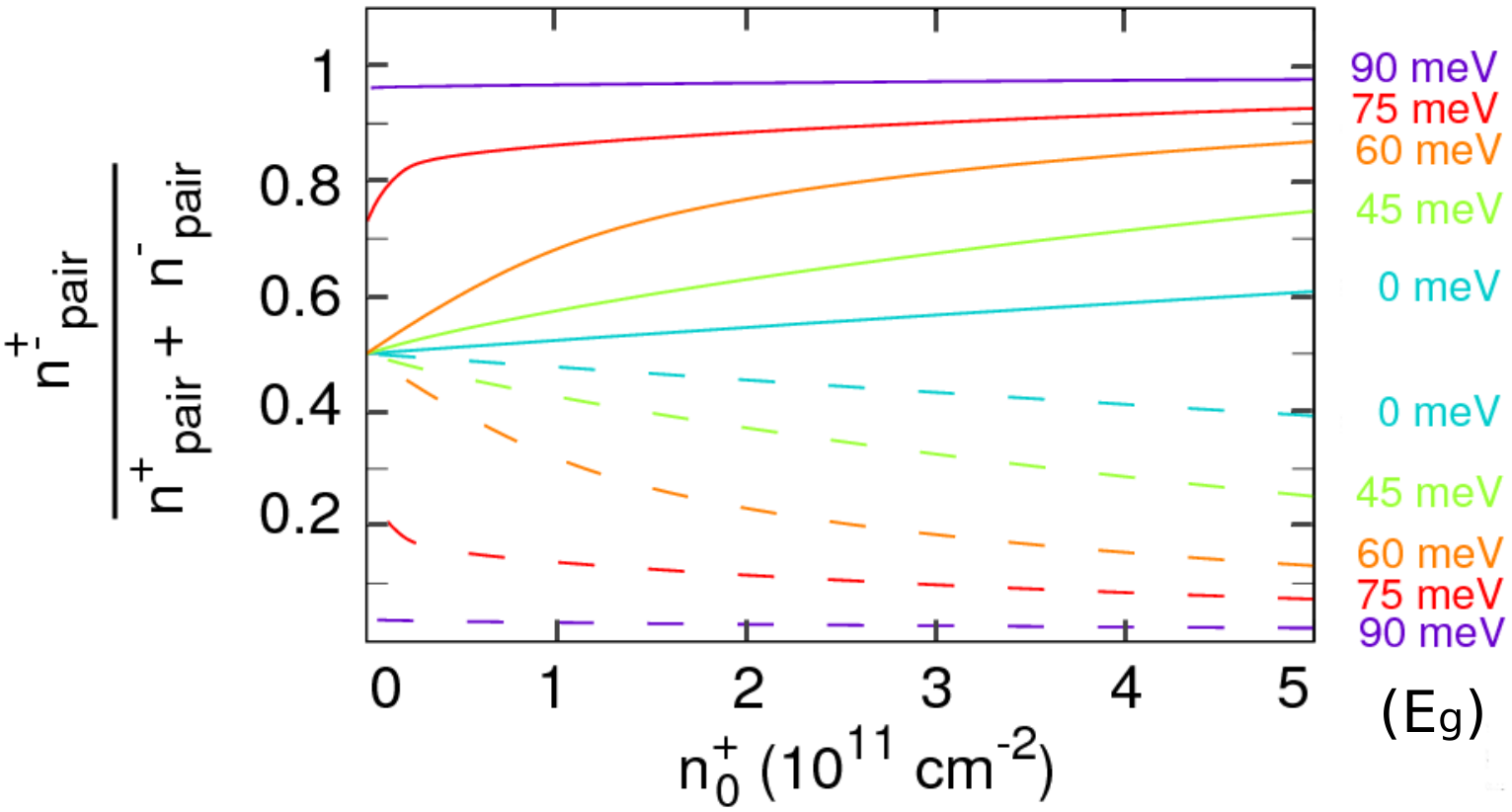}
\caption{Relative number of condensate pairs in conduction band (solid lines) and valence band (dashed lines) as functions of $n^+_0$ for  
different values of the energy band gap $E_g$.}
\label{fig:partial}
\end{figure}

\begin{figure}[t]
\centering
\subfigure
{\includegraphics[width=0.355\columnwidth]{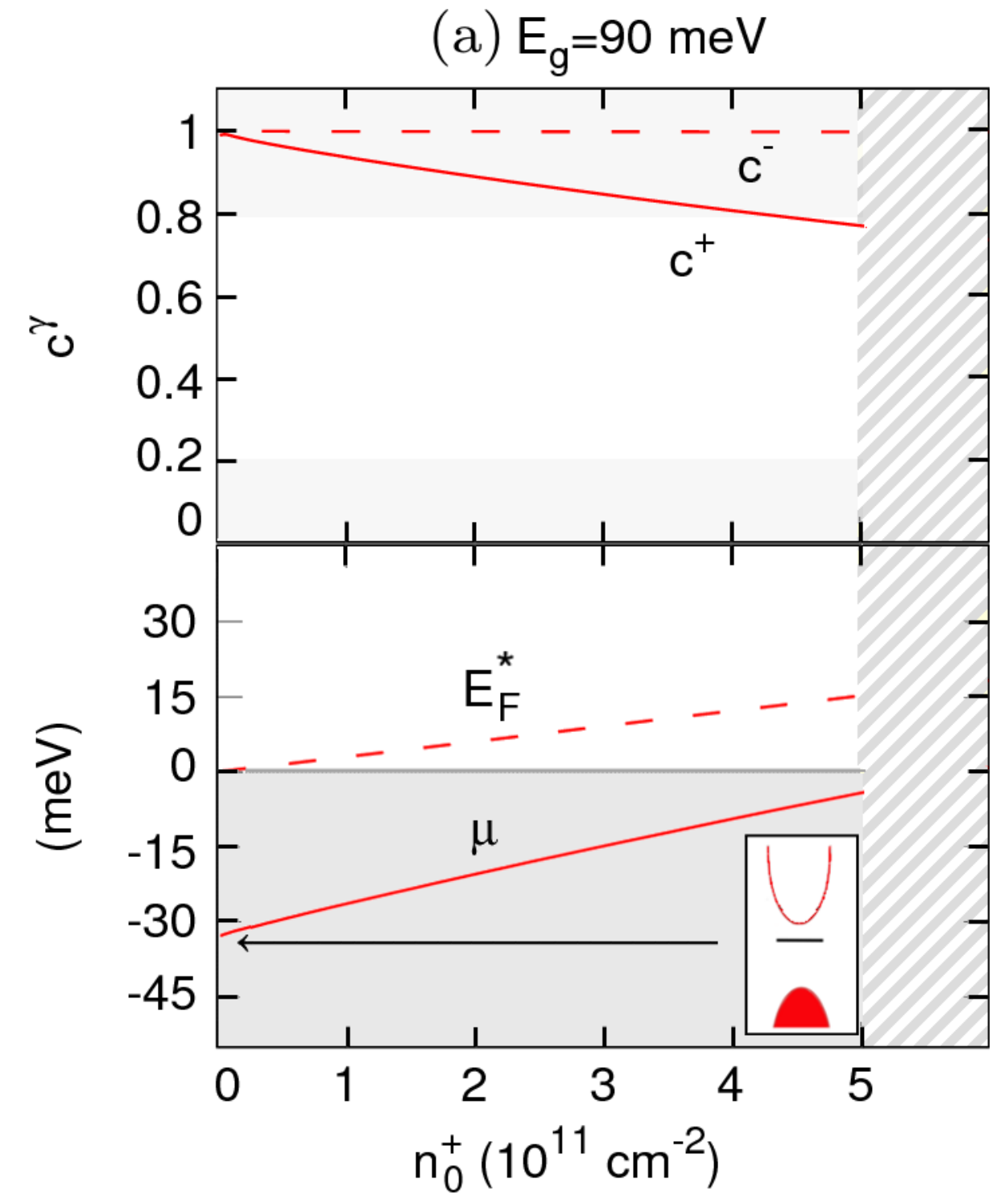} \label{fig:cf90}}
\hspace{-0.32cm}
\subfigure
{\includegraphics[width=0.278\columnwidth]{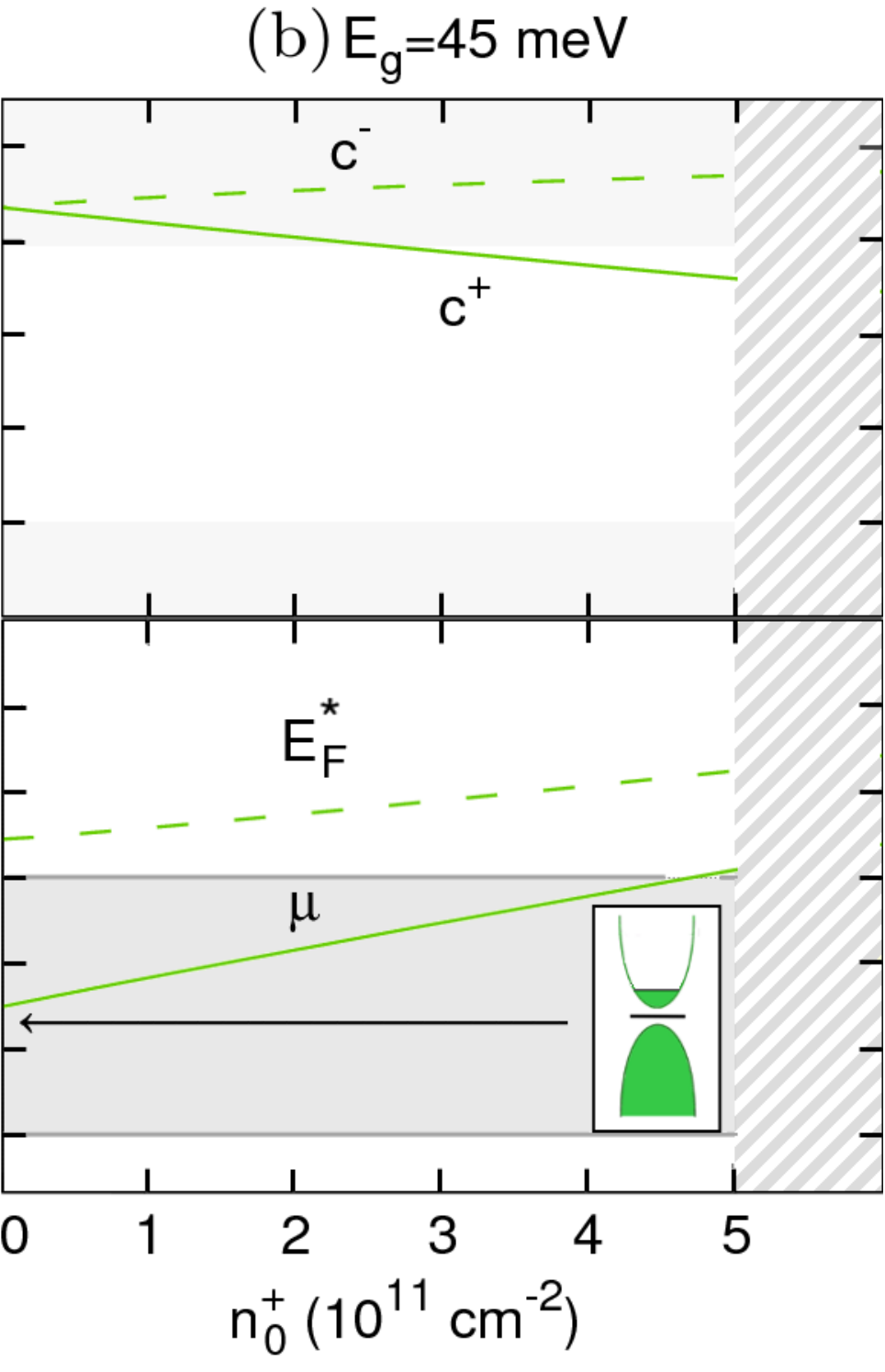}\label{fig:cf45}}
\hspace{-0.21cm}
\subfigure
{\includegraphics[width=0.281\columnwidth]{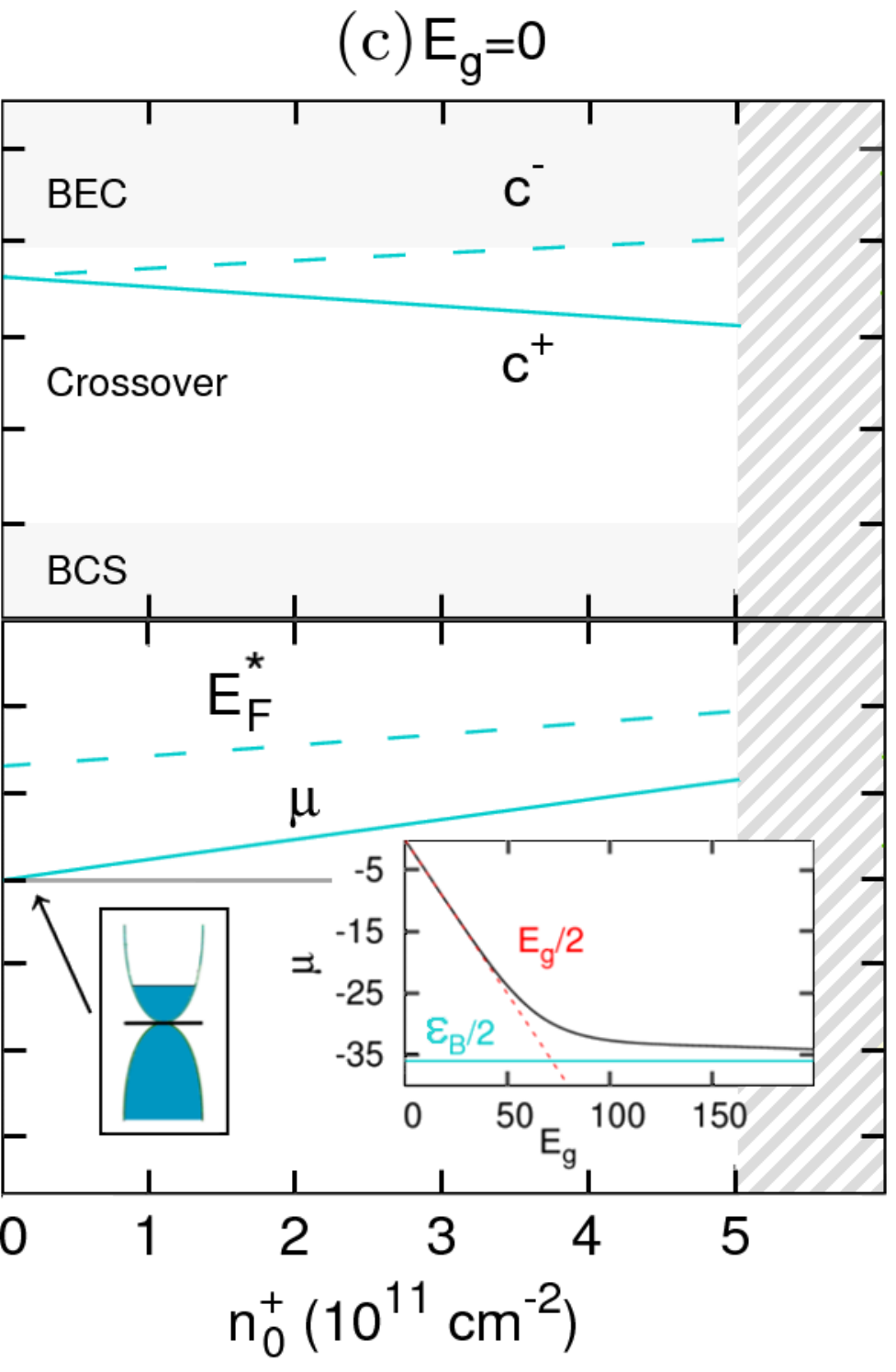}\label{fig:cf0}}
\caption{The condensate fraction and the chemical potential as functions of
$n^+_0$ for different values of $E_g$, as labeled. In the upper panels,
the solid and dashed lines indicate the condensate fraction in the
conduction  and valence band, respectively. In the lower panels, the
solid lines show the chemical potential $\mu$ and the dashed lines the
effective Fermi energy $E_F^*$. The light shaded area represents the energy
band gap. Screening is expected to suppress the superfluidity for
$n^+_0>5.0 \times 10^{11}$ cm$^{-2}$.  Inset shows the limiting value of
$\mu$ as a function of $E_g$.}
\label{fig:condfr}
\end{figure}

Figure\ \ref{fig:partial} compares the contributions to pair formation from the conduction and valence bands as a function of $ n^+_0$.  
The ratios $n^{\pm}_{pair}/(n^+_{pair}+n^-_{pair})$ are shown for different energy band gaps, where $n^\gamma_{pair}=\sum_{k} (u^\gamma_k)^2 (v^\gamma_k)^2$.  
As expected, for large values of $E_g$, pair formation is confined to the conduction band and is independent of $ n^+_0$. 
However for smaller $E_g$, the ratios depend on $ n^+_0$.   We recall that large $ n^+_0$ signifies a small valence band contribution because it contains few anti-particles,
while at small $ n^+_0$ both bands contribute equally to the pair formation, whether the interactions are strong or weak.

Figure \ref{fig:condfr} shows the condensate fractions and the chemical potential as functions of $n^+_0$ for different $E_g$.  
Figure \ref{fig:cf90} is for a large energy band gap, $E_g = 90$ meV, and the behavior of the condensate fraction and  chemical potential 
is indeed close to results for a one-band system \cite{strinati_survey}.  
Using the condensate fraction criterion, for large $n^+_0 \sim 5 \times 10^{11}$ cm$^{-2}$, the conduction band condensate is already in the crossover regime.  
We recall that at values $n^+_0 > 5 \times 10^{11}$ cm$^{-2}$ screening is expected to suppress superfluidity in what would otherwise be the BCS regime. 
As $n^+_0$ decreases, the conduction band condensate eventually enters the BEC regime.  
The chemical potential is less than $E^*_F$ and it becomes negative at the crossover to the BEC boundary.  
For $n^+_0$ going to zero, the conduction band condensate enters the deep BEC limit and $\mu\sim-\varepsilon_B/2$, 
where $\varepsilon_B$ is the  binding energy of an independent electron-hole pair.   
In notable contrast, Fig.\ \ref{fig:cf90} shows that the valence band condensate is trapped in the BEC regime over the full range of $n_0^+$ shown.   
This is because there are very few anti-particles in the valence band when $E_g$ is large.   

Figure \ref{fig:cf45} is for a smaller gap than Fig.\ \ref{fig:cf90}, $E_g = 45$ meV, and the conduction band condensate is slower to enter the BEC regime as $n^+_0$ decreases.  
This is
because  excitations from the valence band now significantly increase the total population of carriers in the conduction band. 
The chemical potential $\mu$ therefore goes negative only at very low $n^+_0$.   
It is interesting that in the zero $n^+_0$ limit, $\mu$ now approaches the mid-point of the energy band gap, $\mu\rightarrow -E_g/2$ 
instead of $-\varepsilon_B/2$,  behavior analogous to  the low  density limit in  a conventional semiconductor.

Figure \ref{fig:cf0} is for $E_g=0$.  In this case there are many carriers in the conduction band excited from the valence band. 
This makes the effective Fermi energy $E^*_F$ significantly larger than in Fig.\ \ref{fig:cf45} at the same $n^+_0$. 
For this reason, the conduction band condensate remains in the crossover regime even for very small $n^+_0$, and the chemical potential $\mu$ remains positive. 
An interesting point is that for a gapless system a negative value of the chemical potential would signify only an inversion of the carrier populations in the bands, so that even for negative values of $\mu$, the system would remain in the crossover regime.

\begin{figure}[t]
\subfigure
{\includegraphics[width=0.355\columnwidth]{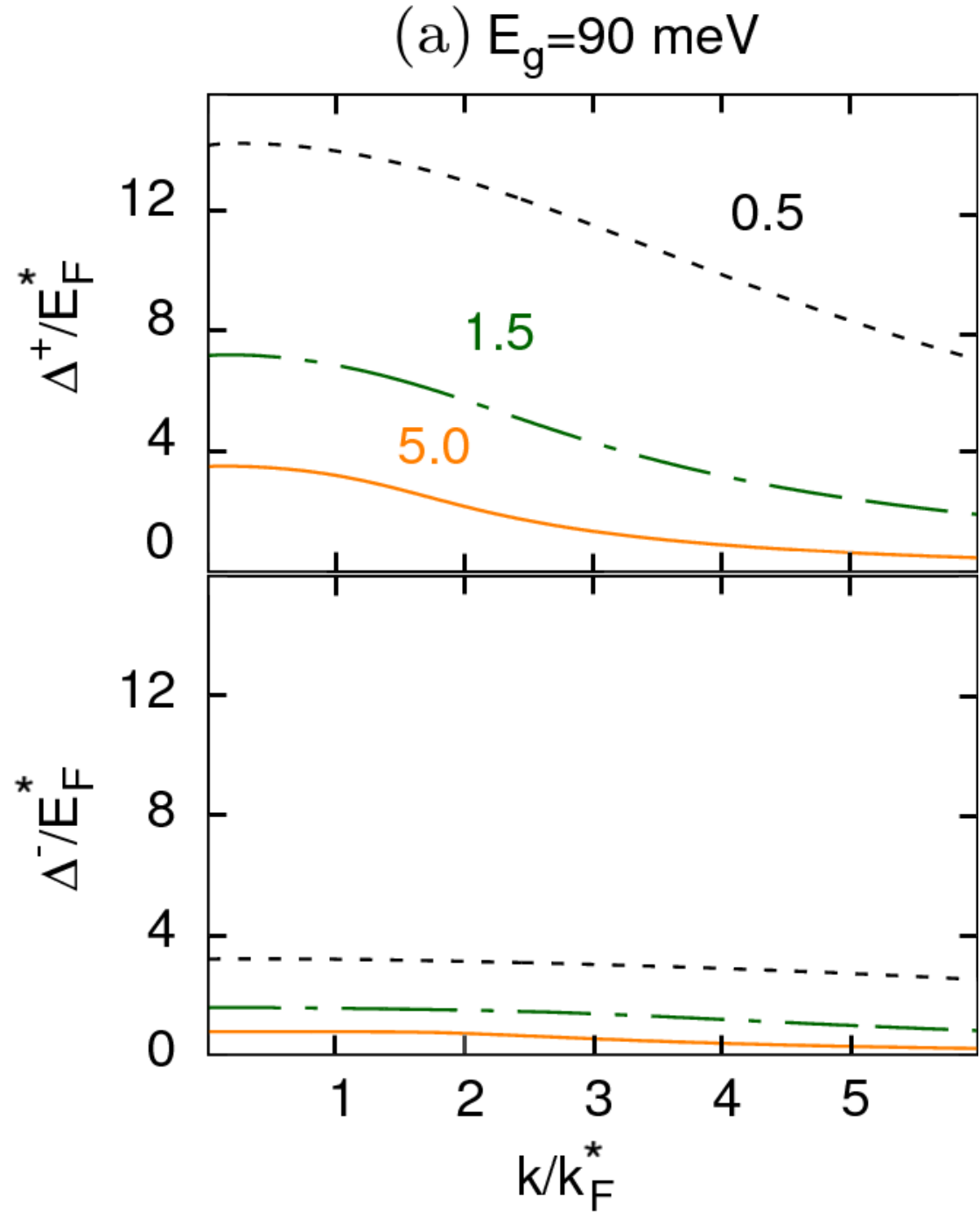}\label{fig:d90}}
\hspace{-0.23cm}
\subfigure
{\includegraphics[width=0.28\columnwidth,keepaspectratio]{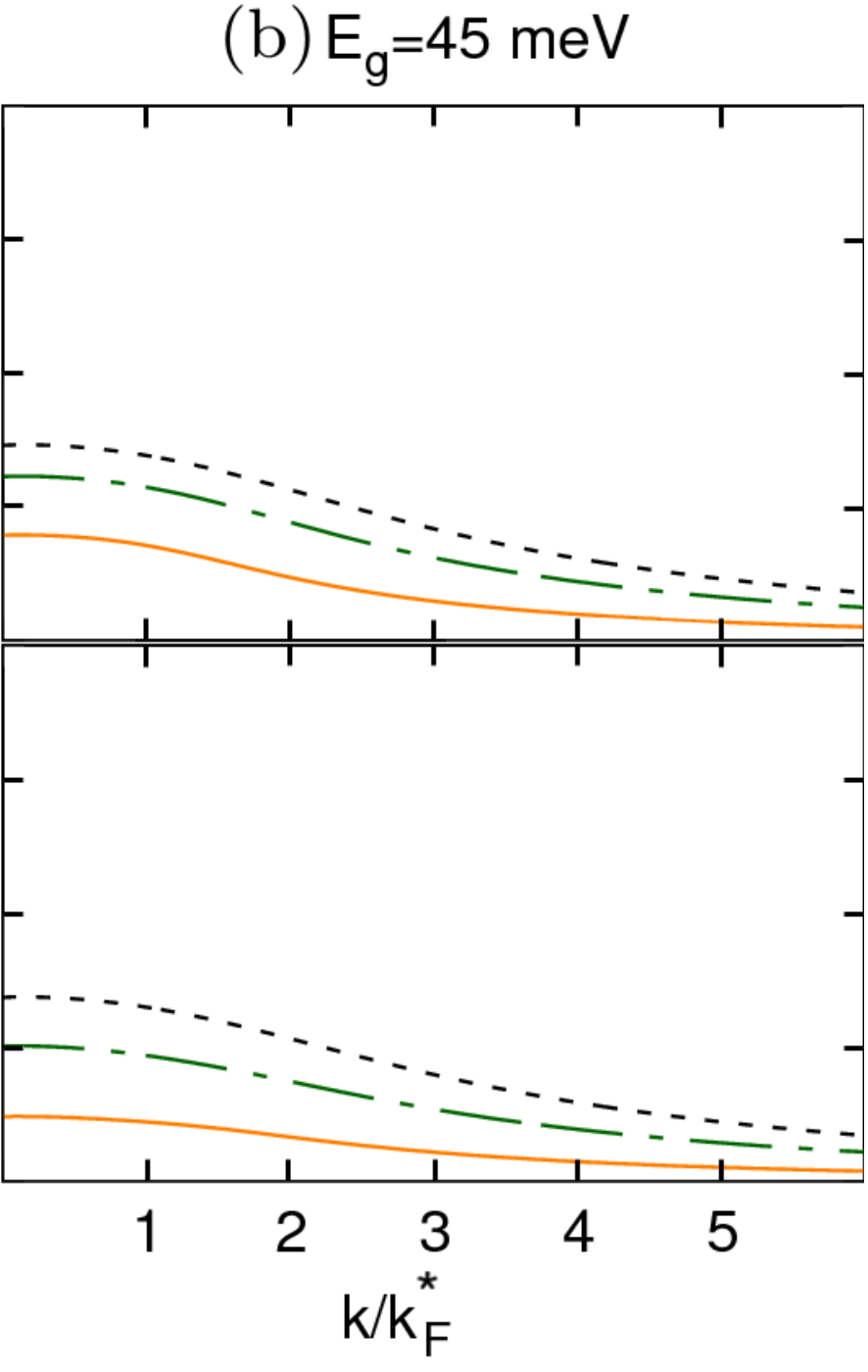}\label{fig:d45}}
\hspace{-0.23cm}
\subfigure
{\includegraphics[width=0.28\columnwidth,keepaspectratio]{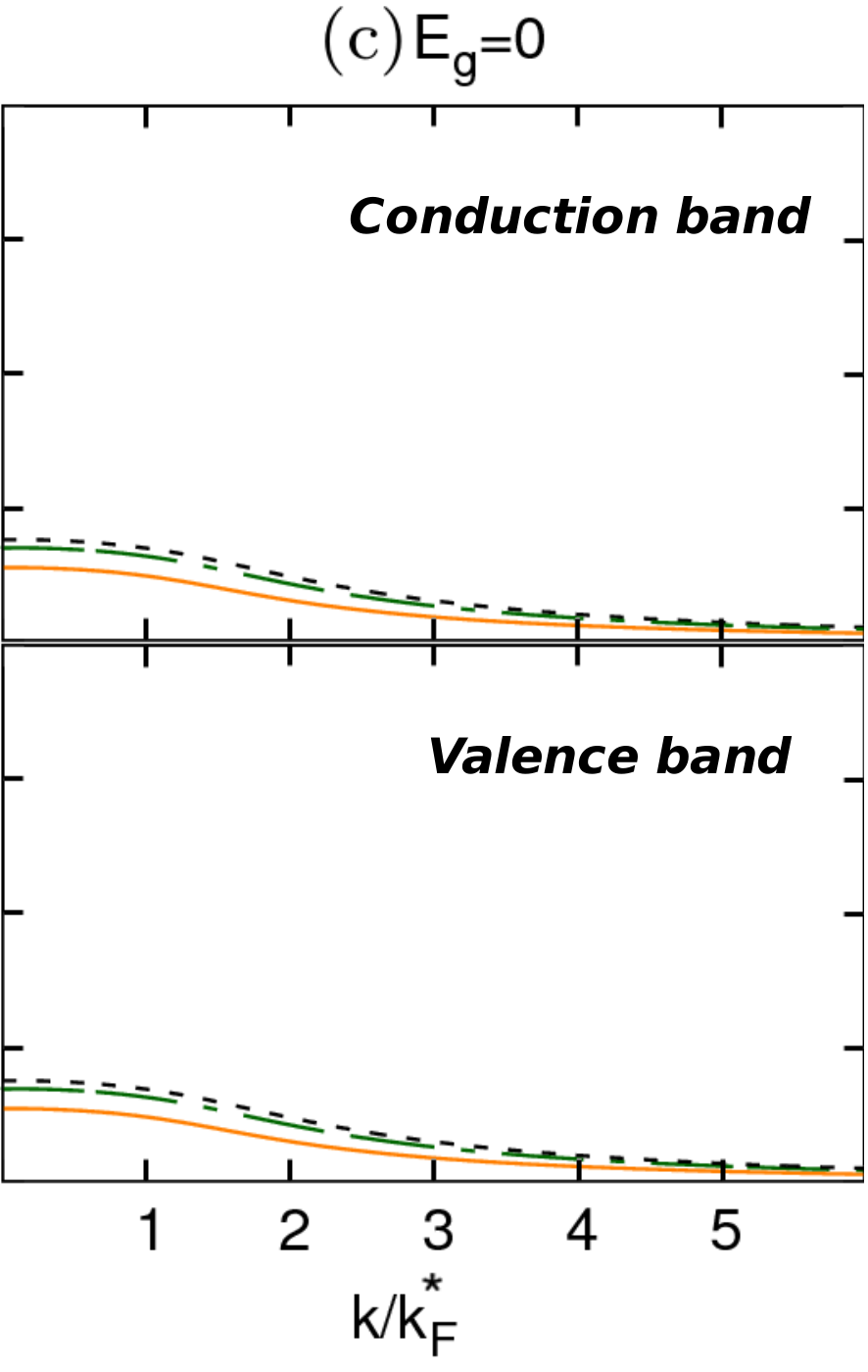}\label{fig:d0}}
\caption{The superfluid gap energy $\Delta^{\gamma=\pm}_k$ in the
conduction and valence bands, for different values of energy band gap
$E_g$.
The results are for low density $n^+_0=0.5 \times 10^{11}$ cm$^{-2}$
(dotted lines), intermediate density $n^+_0=1.5 \times 10^{11}$ cm$^{-2}$
(dashed lines),  and high density $n^+_0=5 \times 10^{11}$ cm$^{-2}$ (solid
lines). }
 \label{fig:delta}
\end{figure}

Figure \ref{fig:delta} shows the momentum dependent superfluid energy gaps $\Delta^{\pm}_k$. 
For large $E_g$, Fig.\ \ref{fig:d90}, the gap equations (Eq.\ \ref{eq:gap}) are nearly decoupled because the $E_g$ term in the form factor 
(Eq.\ \ref{eq:formfactor}) suppresses  $F^{\gamma\gamma'}_{\mathbf{k,k'}}$  for $\gamma\neq\gamma'$.
Then $\Delta^+_k \gg \Delta^-_k$, because of the large energy denominator for $\Delta^-_k$ in Eq.\ \ref{eq:gap}. 
Consistent with the conclusion in Fig.\ \ref{fig:cf90}, the very broad peaks in $\Delta^-_k$ for all $n^+_0$, indicate that the valence band condensate 
for large $E_g$ always remains in the BEC regime. The reason is that the number of paired anti-particles in the valence band, $\sum_k (u^-_k)^2$, remains small for all $n^+_0$.  
For the same reason, the conduction band contains very few carriers excited from the valence band, so the evolution of the conduction band condensate with $n^+_0$,
is very similar to the one-band system: i.e. (i) for small $n^+_0$, $\Delta^+_k \gg E_F$,  its peak is at $k=0$ and it is very broad, characteristics of the BEC regime; 
(ii) for large $n^+_0$, the peak in  $\Delta^+_k $ becomes of order $E^*_F$, it narrows and detaches from $k=0$, though never reaching $k=k^*_F$, 
characteristics of the crossover regime.  

For smaller $E_g$, Fig.\ \ref{fig:d45} and \ref{fig:d0}, the $\Delta^\pm_k$ are comparable and are not very sensitive to $n^+_0$. 
This is because the $F^{\gamma\gamma'}_{\mathbf{kk'}}$ for $\gamma\neq\gamma'$ are no longer small, and so 
strongly couple the two gap equations.
The insensitivity of the superfluid gaps to $n^+_0$  is a consequence of the large number of carriers in the 
conduction band excited from the valence band for all $n^+_0$.  
This means that the total number of carriers in the conduction band remains large for all $n^+_0$.  
Thus the conduction band condensate remains trapped in the crossover regime and is unable to reach the BEC regime even when $n^+_0$ becomes very small.  

Thus, since $F^{\gamma\gamma'}_{\mathbf{kk'}}$ controls the coupling of the two gap equations, 
the dependence of 
$F^{\gamma\gamma'}_{\mathbf{kk'}}$ on $E_g$ for $\gamma\neq\gamma'$ means that 
by tuning $E_g$ we are able, for the first time, to tune the magnitude of the Josephson-like pair transfer.

Our neglect of crosspairing is justified both for large and small gaps $E_g$. 
For large $E_g$  it is clear because of the large energy differences in the corresponding denominators. 
For small $E_g$, the large number of carriers in the conduction band excited from the valence band means a large 
effective Fermi energy, so the crosspairing terms 
again contain large energy difference denominators, reflecting the large energy separation of the carriers in the valence band from the effective Fermi energy.
In addition, the matrix elements for the crosspairing terms are expected to be small 
(see Ref.\ \cite{BreyFertig2007}), and this would further reduce the crosspairing contribution.

\begin{figure}[t]
\includegraphics[width=\columnwidth]{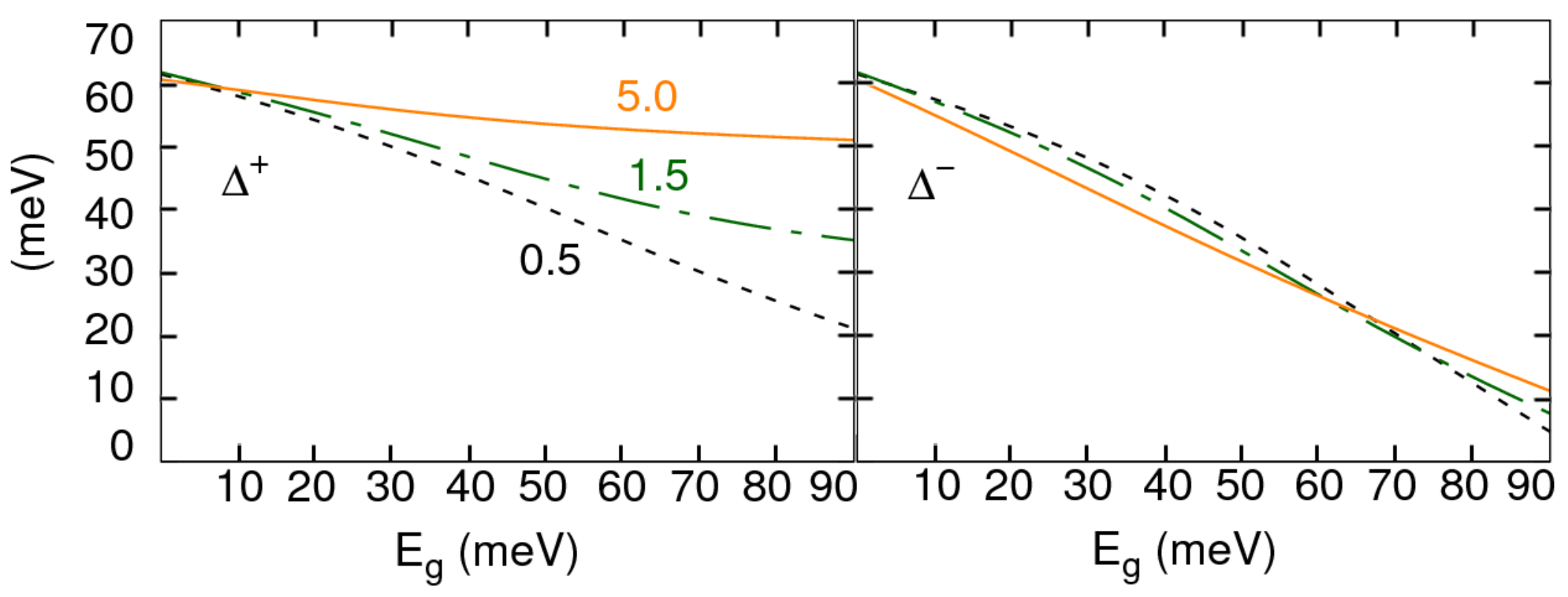}
\caption{The maximum superfluid gap energy $\Delta^{\pm}$ in the conduction and valence bands as functions of the energy band gap $E_g$.
Dotted lines:  $n^+_0=0.5 \times 10^{11}$ cm$^{-2}$; dashed lines: $n^+_0=1.5 \times 10^{11}$ cm$^{-2}$; solid lines: $n^+_0=5 \times 10^{11}$ cm$^{-2}$.}
\label{fig:Delta_meV}
\end{figure}

Figure \ref{fig:Delta_meV} further characterizes the multicomponent nature of the superfluidity.  
As expected, for zero band gap the maximum superfluid gap energy  for the conduction band $\Delta^{+}$ is equal to $\Delta^{-}$,  the maximum gap  for the valence band.
We see in the figure that smaller band gaps, $E_g\alt40$-$60$ meV, significantly boost both $\Delta^{+}$ for the conduction band and  
 $\Delta^{-}$ for the valence band, thanks to the multicomponent property that the contributions from the condensates are additive.  
For too large a band gap, the superfluidity will not be able to take 
advantage of a proximate valence band, and for $E_g\agt90$ meV the valence band condensate is essentially completely decoupled from the conduction band condensate.
This results in $\Delta^+ \gg \Delta^-$, so there is then only one significant 
superfluid gap and one significant condensate.  
Thus, continuously tuning $E_g$ up to higher values will induce, in the same system, a switching-over of the number of superfluid components from
two to one.   However, for optimal conditions for superfluidity, the band gap $E_g$ must also not be too small, otherwise 
excitations from the valence band will maintain  too high a density of carriers in the conduction band.
We recall, as we have discussed, that a high density of carriers inhibits the superfluidity.   


Thus we conclude that a compromise is necessary between selecting too large an $E_g$, which tends to weaken the superfluidity since it excludes the additive 
contributions from the valence band, and too small an $E_g$, which tends to keep the conduction band in the high density regime that is not favorable for superfluidity.
An optimal choice would be in the range $E_g\sim 40$-$60$ meV.
By using the tunable band gap $E_g$, we can move the boundaries of the BCS-BEC crossover while keeping the density fixed.  
When $\sum_k (v_k^+)^2 \sim \sum_k (u_k^-)^2 $ the multicomponent character is evident.  There are two distinct regions: 
(i) For  $E_g\ll E_F^*$, the small $n^+_0$ region remains in the crossover regime even when $n_0^+$ is very small.
The conduction band condensate cannot enter the BEC regime because excitations from the valence band, equal to $g_sg_v\sum_k (u_k^-)^2 $, 
maintain a large number of carriers in the conduction band. 
(ii) When $E_g\agt E_F^*$, the conduction band condensate can enter the BEC regime for small $n^+_0$ because a large $E_g$ suppresses excitations from the valence band. 
These multicomponent properties are reflected in the asymptotic behavior of the chemical potential in the small $n_0^+$ limit.
For large $E_g \geq \varepsilon_B$, the limiting behavior of $\mu$  is the familiar BEC limit $\mu \rightarrow -\varepsilon_B/2$, 
the same as for a single-component system.
However for $E_g<\varepsilon_B$,  it is interesting that the limiting behavior switches smoothly over to the midpoint of the band gap, $\mu \rightarrow -E_g/2$.
This reflects the multicomponent property that for smaller band gaps, $E_g < \varepsilon_B$,  the superfluid is blocked from entering the BEC regime in the low density limit.

\noindent {\it Acknowledgements.} We thank Mohammad Zarenia for useful discussions. 

\bibliography{article}

\end{document}